# The 'Fungibility' of Non-Fungible Tokens: A Quantitative Analysis of ERC-721 Metadata

Sarah Barrington
*School of Information*
*University of California, Berkeley*
Berkeley, CA, USA
sbarrington@berkeley.edu

Nick Merrill
*Center for Long Term Cybersecurity*
*University of California, Berkeley*
Berkeley, CA, USA
ffff@berkeley.edu

## I. ABSTRACT

Non-Fungible Tokens (NFTs), digital certificates of ownership for virtual art, have until recently been traded on a highly lucrative and speculative market. Yet, an emergence of misconceptions, along with a sustained market downtime, are calling the value of NFTs into question. This project (1) describes three properties that any valuable NFT should possess (permanence, immutability and uniqueness), (2) creates a quantitative summary of permanence as an initial criteria, and (3) tests our measures on 6 months of NFTs on the Ethereum blockchain, finding 45% of ERC721 tokens in our corpus do not satisfy this initial criteria. Our work could help buyers and marketplaces identify and warn users against purchasing NFTs that may be overvalued.

## II. INTRODUCTION

On March 11th 2021, American artist Mike Winkelmann (professionally known as Beeple) sold a JPG image for over $69million[1]. Despite having previously never sold a piece for more than $100[2], the auction at Christies lead him to become one of the top three most valuable living artists behind David Hockney and Jeff Koons[3]. Unlike Hockney and Koons' art, Beeple's piece existed digitally as a .PNG image file within a public contract.

Beeple's image, Everydays: The First 5000 Days, is a Non-Fungible Token (NFT), which acts as a digital certificate of ownership for a unique physical or digital asset[4]. Since, the NFT market has grown to become robust, experiencing over $17.6 billion dollars in sales in 2021[5]. Yet, it is also speculative, variously described as irrational and overhyped; with its exponential market movements leading to comparisons being drawn to the 'Tulip Mania' of 17th century Holland[6].

What gives NFTs value? This paper argues that three key properties (permanence, immutability and uniqueness) are necessary (but not sufficient) conditions for an NFT to have fundamental value. Building on this argument, we establish the first quantitative summary of permanence and immutability. We test this metric on 6 months of data gathered from the Ethereum blockchains. From our data, We estimate that 45% of ERC721 tokens exist in non-permanent, centralized storage; as opposed to directly on the Ethereum blockchain. This includes 8% that are stored in one of the top 10 leading cloud providers, including Heroku apps, Amazon S3 storage and Google cloud, which rely on traditional web 2.0 platform infrastructure to remain online. We observe that only 72.43% of ERC721 tokens are readable without specialized decompilation.

## III. VALUING NFTS

NFTs are typically represented as 'smart contracts' on decentralized public ledgers such as the Ethereum Blockchain[7]. Although multiple blockchains and side-chains enable NFT activity, including Solana[8] and Polygon[9], the Ethereum ERC-721 protocol is considered the most widely adopted as it provides a standardized and minimal framework to facilitate the creation of NFTs[10][11].

We thus focus our analysis on the ERC-721 protocol and present three properties inherent to Ethereum NFTs, and as such, three basic criteria for an NFT to possess fundamental value.

**Permanence**
For ERC-721 tokens, a permanent and public record of activity relating to each NFT artwork is available through interacting with the Ethereum blockchain[12][13].

The ERC-721 token enables the storage of NFT details through smart contract *metadata*, a series of fields that can be queried to return specific and unique details about the digital asset. By contrast, cryptocurrencies such as Bitcoin and the Ethereum ERC-20 standard are considered fungible tokens, as their underlying asset metadata is interchangeable[14].

**Immutability**
Through implementation of ERC-721 tokens, an NFTs smart contract and its provenance cannot be modified due to the permanence of blockchain records[15][16][17].

Although the contract is permanently and immutably stored on the ethereum blockchain, the metadata itself can be stored elsewhere- which can lead to conceptual inconsistencies between on and off-chain data[18]. Most crucially, the TokenURI metadata method frequently provides a URI link to an asset, as opposed to containing the set itself.

This means that in practice, although the contract is stored permanently and immutability on a blockchain, the artwork is often stored elsewhere- including in traditional, centralized storage methods. In a study performed by audit site CheckMyNFT, several high-profile NFT artists were found to have created NFTs that lead to errors or non-existent assets[19][20]. Referred to as the 'Broken Link Problem'[21], the storage of images '*off-chain*' presents a potential major weakness in the implementation of NFTs.

**Uniqueness**
Fungibility describes an asset's ability to be interchangeable, in the same way that a physical dollar bill can be interchanged with another and still retain the same value[22]. The absence of fungibility, alongside the presence of permanence and immutability of an NFT, leads to the definition of uniqueness as an implicit property. This fundamentally ascribes value to NFTs; through mirroring the concepts of physical scarcity and authenticity of art in the non-virtual world[23].

Yet, with off-chain storage, the validation of uniqueness of underlying assets is presently challenging. Although we do not explicitly measure this property as an initial criteria in this paper, propose further work in section **VII** outlines methods by which to explore and measure this concept.

## IV. DATA AND METHODOLOGY

**Dataset Selection**
Our primary datasource is a transaction record of on-chain ERC721 NFT activity for a 6 month period, originally created by moonstream.to and downloaded as a SQLite database from online Data Science community Kaggle[24]. The moonstream.to dataset was constructed by tracking all 'transfer' activities emitted by ERC721 tokens on the Ethereum blockchain for this time period[25], and thus is assumed to represent a comprehensive list of all NFTs in circulation between these dates.

This dataset contains 9,292 unique addresses for a 6 month period in 2021, comprising of 7,020,950 unique tokens circulated between 1st April to 25th September 2021. The 'mint' (creation) and 'transfer' activity tables were combined, with duplicated actions dropped, to create a single reference list of all ERC721 NFTs in circulation between these dates.

Several other sources were explored; including a CSV export of the NFT token tracker from leading ethereum blockchain explorer etherscan.io[26], open source academic datasets such as from *Mapping the NFT revolution: market trends, trade networks, and visual features, Nadini et. al*[27] and directly from the ethereum blockchain itself. Each was constrained due to API and computational limits, or pulling only from a limited selection of marketplaces and popular tokens.

**Smart Contract Execution**
Obtaining metadata for each token presents several complexities. We thus developed a framework for obtaining the URIs of each token through a three step process. Firstly, we use the primary reference list of NFTs as a source for all NFT activity within a finite time period.

Secondly, we extract the ABI (Application Binary Interface) from the contract using the etherscan.io ABI endpoint, in order to obtain the contract's readable and executable source code. Finally, the contract address, alongside its ABI, are passed to an Ethereum MainNet node, that allows the tokenURI method to be executed for a specific token ID associated with that contract.

**Dataset Assumptions**
Within the primary reference dataset, several 'collections' (singular smart contracts that cover multiple tokens) contained thousands of tokens, it was noticed that executing each tokenID individually returned the same root URL. In order to speed up computation on the Ethereum node, we thus assume that each 'collection' (contract) uses the same storage mechanism for their tokens.

We validated this assumption using a sample of 30001 NFT transactions for a 45hour period on 1st and 2nd April 2021. These represented 21186 tokenIDs with readable contracts, which in turn, represented 188 unique contract addresses. Each address returned exactly one URI stem- most often comprising the smart contract address and given TokenID, for example: 'IPFS.io//contractaddress_tokenid'.

## V. FINDINGS

**1. Out of 9292 ERC721 sample collections (contracts), only 72.43% have human-readable source code**
Contract source code exists on the Ethereum blockchain as 'bytecode' only. Smart contract developers can provide an Application Binary Interface in order to allow the contract source code to be retrieved as a decompiled and readable JSON file. Third party block explorers, such as etherscan.io, can provide API services to extract these ABIs in order to execute the source code. Fig. 1 shows the readability by category for all contracts provided to the etherscan.io API, for which 72.43% of contracts were deemed human or machine readable without specialized decompilation.

After obtaining the reference list of all NFT activity, the unique addresses were extracted and the etherscan 'contractABI' API endpoint was used to obtain the corresponding ABI as a JSON file. Errors were returned if a contract did not return an ABI response; signaling that the contract exists only as raw bytecode that has yet to be decompiled by etherscan, or if a given TokenID failed at the TokenURI method.

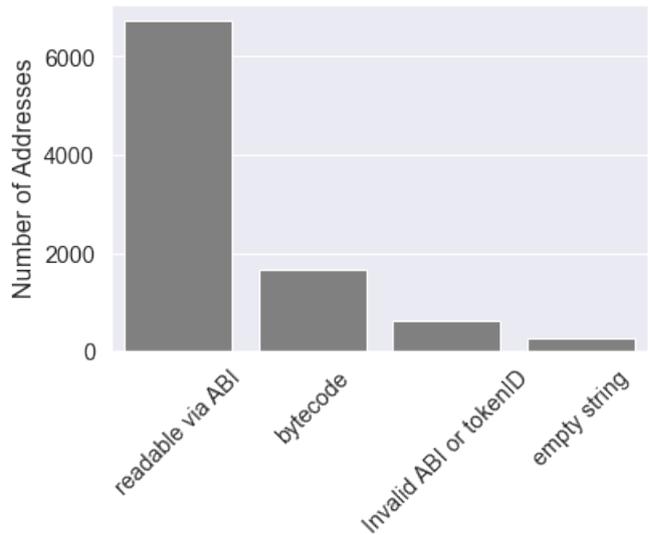

Fig. 1: Readability by category for all contracts provided to the etherscan.io API

**Error Logs Analysis**
There were two points of failure recorded for the TokenURI retrieval process as follows:
- Failure to obtain the contract ABI
- Failure to obtain a valid TokenURI for a given contract ABI

Through code design, different values were recorded for each error- recording a contract read failure as well as tokenURL failures. Through analysis of error logs, 15 example addresses were extracted at random in order to validate the content of the error messages. Of the 9,200 unique contract addresses, 72.4% provided interpretable ABIs with valid tokenURI methods. Examples of error messages can be found in the appendix (**App. 1**). The remaining addresses can be categorized as follows:

**i. Bytecode only**
In 17.97% of contracts, a readable ABI has not been provided alongside the contract, and the raw byte code has not been manually decompiled by the Etherscan block explorer. Decompilation can be performed if requested from a provider, but there is no guarantee that a provider will service all requests (see sections **VI** and **VII**).

**ii. 'Invalid ABI or TokenID'/source code**

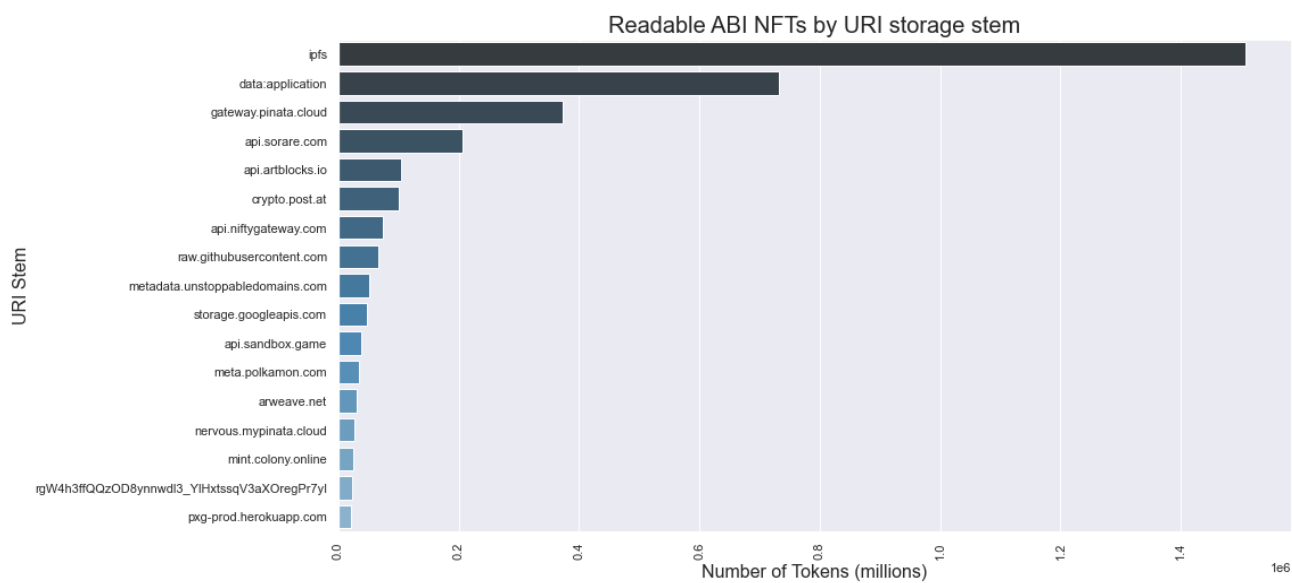

Fig. 2: The 17 highest frequency URI stems of token locations

This failure arose in 6.78% of contracts when tested using a single tokenID from the reference list, and an invalid response was returned. This failure was found to be the result of two root causes:

a) An invalid tokenID was provided - despite being a valid contract with a readable ABI containing the TokenURI method, the recorded TokenID returned a null response.
b) The TokenURI method doesn't exist for that contract - it is possible that asset data could be contained within a different metadata field. However, it was assumed that a failure on this method meant that most likely the NFT did not contain any digital asset in the form of an image or video.

**Iii. Empty string**
In 2.82% of cases, a successful query to a valid contract with a readable ABI returned an empty string. Upon validating this with the etherscan.io online search tool, it was concluded that these contracts intentionally return an empty TokenURI string.

**2. 45.06% of the sample ERC721 NFTs (tokens) metadata are stored in non-permanent, centralized and mutable storage**
Having obtained an example TokenURI for each smart contract, the hosting root was extracted alongside the host protocol through splitting the 'tokenURI' field according to the backslash character. This enabled the separation of URI root and contract or asset specific information (e.g. splitting 'https://blockdatanalysis.com/v1/22' to become 'https' and 'blockdataanalysis.com').

The hosting root and protocol were then duplicated across all tokens in a given collection, using the earlier assumption that each tokenID within the same smart contract (collection) would use the same storage mechanism.

**URI Analysis**
The host roots were extracted, and the top 17 URI stems by number of tokens are shown in Fig. 2. The roots of these URIs were then grouped according to 6 overarching storage categories.

**Permanent: Decentralized**
1. IPFS - InterPlanetary File Storage, the leading decentralized storage solution that, similarly to the Ethereum blockchain, provides a peer-to-peer network supporting a distributed file system[28]. Although IPFS still relies on network hosts, control is returned from marketplaces and sellers to buyers as buyers can pay financial contributions to keep their NFT present on the network[29].
2. Piñata - a provider of customizable dedicated gateways for accessing the IPFS network, without the rate-limiting associated with public IPFS gateways[30].

**Non-Permanent**

3. Cloud storage - URI locations using known cloud storage providers, with particular focus on the top 10 cloud providers of 2021[31]. The regular expression for extracting these URIs included the following: *['heroku', 'aws', 'google', 's3', 'azure', 'microsoft', 'kyndryl', 'digitalocean', 'linode', 'alibaba', 'oracle', 'tencent']*.

4. Private domain or other - any other domain not contained in any other category, often taking the form of 'https://....com'

**On-Chain**

5. Base64 encoded JSON - all TokenURIs containing a raw JSON string encoded in base64, distinguishable through beginning 'data:application/json;base64,'. These files require two layers of decompiling, firstly for the tokenURI metadata, and secondly for the SVG image itself beginning 'data:image/svg+xml;base64'. Because all information about the asset is contained within the returned string, these NFTs can be classified as stored **on-chain,** and are truly permanent according to the underlying principles of the Ethereum blockchain they are stored directly on.

**Unreadable source code**

6. Invalid contract or tokenID - from the 27.57% non-readable contracts (comprising 14.75% of total token volume). These are included in all percentages for completeness of analysis.

A breakdown of token numbers by storage category can be found in the appendix (**App. 2**). In aggregating the above 6 categories to permanent, non-permanent and not readable, the results in Fig. 3 are observed.

| permanence | n_tokens | % |
|---|---|---|
| non-permanent | 3163820 | 45.07 |
| not readable | 1035763 | 14.75 |
| permanent | 2821367 | 40.18 |
| Total | 7020950 | 100 |

Fig. 3: Summary of asset storage by permanence category

## IV. DISCUSSION

The results of the TokenURI initial analysis indicate that in most cases, two core elements of the NFT value proposition- permanence and immutability- apply only to the smart contract as opposed to the underlying artwork itself. In this section, we explore baseline technical extensions we could implement to further validate these findings, before examining how these results lead to bigger questions around uniqueness and the need for a formal validation system.

**Dataset & technical limitations**

The mainstream.to dataset is assumed to contain a comprehensive list of all tokens in circulation for the 6 month period during 2021. Should this be proven not to be the case, the percentages provided in the above analysis will need to be revalidated against the updated full list of NFTs in circulation for this time period. Additionally, the circulation has likely changed since 2022, and so work is in progress to obtain a comprehensive new dataset containing all known tokens in circulation.

There are also options to extend the technical pipeline to include the decompilation of byte code, which presently requires the manual use of third party tools such as the Online Solidity Decompiler from ethervm.io[32]. Additionally, further examination of the 'private domain or other' category could reveal more granular information about how private domains store token metadata; and in particular, how reliant these domains are on 3rd party hosting requirements (for example, in failing to pay the website hosting fees, could an NFT's metadata become non-existent?).

Finally, in many cases, the imageURL is not returned directly by the 'tokenURI' method, but is included as part of a JSON response and thus requires additional steps to view the asset. An analysis of 'click depths' between the tokenURI method and the final asset view will be explored for a further study.

## VI. FURTHER WORK

**Examining Uniqueness**

We propose that there are three separate levels of uniqueness to explore. Firstly, the uniqueness of the token ledger entry on the blockchain. This can be assumed to be unique through the inherent infrastructure of the Ethereum blockchain, in which each new transaction belongs to a block identified by a distinct hash, generated using the SHA256 cryptographic hash algorithm[33].

Secondly, the uniqueness of token metadata that is returned via the TokenURI method. Although the ledger entry is unique, presently there is no validation as to whether the metadata returned by the ledger entry and contract is unique- for example, if two tokens returned different tokenURI methods, but ultimately the same metadata was found in both locations. This can be evaluated through a click depth analysis of all tokenURI responses, in which URIs are followed and evaluated at each stage until the end asset is found.

Thirdly, and most importantly, the uniqueness of the underlying asset itself. Presently, it is difficult to evaluate token uniqueness, and buyers are offered few assurances that can prove the asset is genuinely permanent, unique and in existence. 'Rug-pull' and wire fraud such as the cases of *Frosties*[34] and *Finiko*, in which buyers invest financially into a cryptocurrency or NFT project only to find the project shut down soon after, increased by 81% from 2020 to 2021, with total estimated damages of $7.7bn[35]. Similarly, between July 2021-22, an estimated $100million worth of NFTs were stolen through fraudulent activity[36][37]. In oder to protect against these kinds of threats, the location and integrity of the underlying asset must be known and understood by the owner.

Additionally, decentralized storage may provide permanence so long as the network is maintained, but presently offers no methods for assessing whether the underlying asset is unique (even if hosted on a unique network location).

Image hashing provides one such methodology for validating the uniqueness of an image or video. In particular, perceptual hashing generates a unique 'fingerprint' of a multimedia file[38] that can then be matched to future uploads in order to detect previously identified content[39]. PhotoDNA[40], developed by Professor Hany Farid and Microsoft Research, is an example of one such technology that performs this operation for the detection and removal of online Child Sexual Abuse Material.

**The need for a validation system**
The findings from this initial research indicate the need for a validation system in which buyers can be assured of the quality of an underlying digital asset.

In future work, we will propose a validation system in which tokens can be queried against a combination of the metrics discussed above, including storage mechanism of the asset metadata, click depth of TokenURI to asset and scarcity of image hash. In turn, this system will generate a score which can be re-associated with given buyers, sellers and marketplaces to assist the verification of all stakeholder involved in the NFT activities.

NFTs encompass the fundamental technologies required to digitize the world of art, and furthermore, revolutionize how the concepts of physical scarcity and ownership are represented online. We propose that if this system were widely deployed, we could give assurances to users of marketplaces, evaluate the marketplaces themselves, and increase the overall reliability and transparency of the NFT ecosystem in order to realize this potential.

# APPENDIX

| Error | Address | TokenID | Etherscan API response |
|---|---|---|---|
| Raw Bytecode only | 0x5033973ea65C66A8745acDB4f8ecb326365de2Be | All | ERROR __mp_main__ Contract source code not verified -- NOTOK |
| Invalid tokenID | 0x000000000437b3CCE2530936156388Bff5578FC3 | 1 | error': {'code': 3, 'message': 'execution reverted: ERC721Metadata: URI query for nonexistent token' |
| Invalid TokenURI method | 0x017bBa5d5D32feb687FDAfB9700418d55dAad091 | 454 | ERROR __mp_main__ ("The function 'tokenURI' was not found in this contract's abi. ", 'Are you sure you provided the correct contract abi?') |
| Empty string | 0x0000009FC3Fea00F2e750632d49E2AfD96878F2a | 1 | 'string : ' |

App. 1: Example error responses for contracts of each readability category

| Storage type | | Number of tokens | % | Example contract | tokenID | tokenURI metadata | End image URI |
|---|---|---|---|---|---|---|---|
| Non-permanent (centralized) | 3rd party cloud provider (Amazon AWS) | 501489 | 7.14 | 0x0c4bAE424DEAA9Be96bb0998524Bc91e1903D794 | 1652610435 | https://pellar-dev.s3-ap-southeast-1.amazonaws.com/nft/1652610435.json | https://pellar-dev.s3-ap-southeast-1.amazonaws.com/nft/images/nft/images/1652610435.jpg |
| | Private domain or server | 2662331 | 37.92 | 0x00CCc5Fe33fa66847082af413d4A8700cd7CDe16 | 0 | https://www.pullmyrug.com/api/metadata/0/ | http://www.pullmyrug.com/api/image/0/ |
| Permanent (Decentralized) | IPFS | 1515966 | 21.59 | 0x0025Eae58dF9F636F261CFdFa98cAcb57779DF74 | 10 | https://ipfs.io/ipfs/QmeW27ViBBpJWo9mDqg9Bpq9KLbHiFGAE9Qrzs7TyGMwvi | ipfs://ipfs/QmSS78vR7kforvUU9jk9JjSowc1GDtGDkg2mktAcWtPWT3/image.png |
| | Piñata | 558730 | 7.97 | 0x06bCa1e513603a5544E0A70256607087ABa73659 | 2 | https://gateway.pinata.cloud/ipfs/QmbuE31SxEDjfVrK26pH1ktdhMkf42WLXcMTrVWwGzSVcK/748 | https://gateway.pinata.cloud/ipfs/QmQYHo6VeSAXUWVZTDpqXLjyJvKjJLABiRAYE1kxRRuUAB/748.png |
| Permanent (On-chain) | Base64 encoded JSON + SVG | 746671 | 10.63 | 0x05a46f1E545526FB803FF974C790aCeA34D1f2D6 | 10 | data:application/json;base64,eyJuYW1lIjogIk4gIzEw… | data:image/svg+xml;base64,PHN2ZyB4bW… |
| Invalid ABI or TokenID | | 1035763 | 14.75 | N/A | N/A | N/A | N/A |
| Total | | 7020950 | 100 | | | | |

App. 2: Example contracts and URIs for each storage type